\documentclass[preprint]{JHEP3} % 10pt is ignored!

\usepackage{graphicx}
\usepackage{epsfig}
\usepackage{psfig}
\usepackage{bm}
\def\gtap{\mathrel{ \rlap{\raise 0.511ex \hbox{$>$}}{\lower 0.511ex
   \hbox{$\sim$}}}} \def\ltap{\mathrel{ \rlap{\raise 0.511ex
   \hbox{$<$}}{\lower 0.511ex \hbox{$\sim$}}}} 
\newcommand{\beq}{\begin{equation}}

\newcommand{\eeq}{\end{equation}}
\newcommand{\bea}{\begin{eqnarray}}
\newcommand{\eea}{\end{eqnarray}}
\newcommand{\lsim}{\stackrel{<}{\scriptstyle \sim}}

\newcommand{\MeV}{\mbox{$ \ \mathrm{MeV}$}}

\newcommand{\nus}{\nu_h}
\newcommand{\umu}{\mbox{$|U_{\mu h}|^2$}}

\title{ Bounds on heavy sterile neutrinos revisited}

\author{
Alexander Kusenko\\ Department of Physics and
 Astronomy, UCLA, Los Angeles, CA 90095-1547\\
 \email{kusenko@ucla.edu}}
\author{Silvia Pascoli\\ Department of Physics, Theory Division, CERN, CH-1211 Geneva 23, Switzerland \\IPPP, Department of Physics, University of Durham, Durham
  DH1 3LE, United Kingdom\\
\email{Silvia.Pascoli@cern.ch}}
\author{Dmitry Semikoz \\PC, College de France, 11 pl. Marcelin Berthelot,
Paris 75005, France\\INR RAS, 60th October Anniversary prospect 7a, 117312
Moscow, Russia\\
\email{dmitri.semikoz@cdf.in2p3.fr}}

\abstract{ We revise the bounds on heavy sterile neutrinos, especially in
the case of their mixing with muon neutrinos in the charged current.  We
summarize the present experimental limits, and we reanalyze the existing
data from the accelerator neutrino experiments and from Super-Kamiokande to set
new bounds on a heavy sterile neutrino in the range of masses from 8~MeV to
390~MeV.  We also discuss how the future accelerator neutrino experiments can
improve the present limits.  }

\begin{document}
\section{Introduction}

The existence of three active neutrinos and of the mass eigenstates
$\nu_1,\nu_2,\nu_3$ is well established, but the existence of a forth heavy
mass eigenstate, $\nu_h$, mainly in the direction of a sterile neutrino,
remains an open question.  Sterile neutrinos are
SU(3)$\times$SU(2)$\times$U(1) singlet fermions, which can mix with
ordinary neutrinos. 
 
There many reasons why one is interested in the limits on sterile neutrinos.  
Heavy mostly-sterile neutrinos have been investigated
for their role in cosmology and astrophysics (see, e.g.,
Ref.~\cite{Dolgovrept}).  A keV sterile neutrino is a viable dark matter
candidate \cite{WDM,lowTr}, which can also explain the origin of the pulsar
kicks~\cite{pulsarkicks}.  Sterile neutrinos provide a viable framework for 
baryogenesis~\cite{akh,Asaka:2005pn}. 
It was pointed out~\cite{511line} that if such heavy
neutrinos constitute a small but non-negligible fraction of dark matter,
their decays into $e^+$-$e^-$ pairs might produce the 511 keV gamma line
observed by the INTEGRAL $\gamma$-ray observatory.  Decays of heavier
neutrinos have been proposed to explain the early ionization of the
Universe~\cite{reioniz}.  If neutrinos are Majorana particles, they could
mediate processes that violate the lepton number by two units, such as
neutrinoless double beta decay, muon-positron conversion, and rare kaon
decays~\cite{dl2}.  In particular, for masses $245 \MeV <m_h<388 \MeV$, the
decay $K^+ \rightarrow \pi^- \mu^+ \mu^+$ could be strongly
enhanced~\cite{dib}.  From the theoretical point of view, heavy sterile
neutrinos with masses in this range arise naturally in extended technicolor
models~\cite{Piai}.  Because of these interesting possibilities, we want 
to map out the parameter space available for sterile neutrino masses and 
mixing parameters.

Here we consider the bounds on heavy neutrinos
with masses $1 \MeV \ltap m_h \ltap 400 \MeV$, which are produced in pion,
muon and kaon decays.  
We will review the existing limits on the mass and mixing of a heavy,
mostly-sterile,
neutrino~\cite{Shrock81,prodpion1,prodpion2,prodpion3,prodpion4,prodpion5,prodpion6,prodpion7,prodpion8,prodpion9,Asano81,Hayano82,Berg83,Bernardi86,Bernardi88,Bar93}. 
We will then derive new bounds based on a re-analyses of the data from neutrino
oscillations and accelerator neutrino experiments.  In the 10--390~MeV mass
range we use the data from accelerator
experiments~\cite{Berg83,Bernardi86,Bernardi88,Bar93}.  For masses $8 \
\MeV \ltap m_h \ltap 105 \ \MeV$, we use the Super-Kamiokande (Super-K)
data~\cite{SK2003}.  We will also discuss ways in which the present and
future neutrino oscillation experiments, such as MiniBOONE~\cite{miniboone},
K2K~\cite{k2k} and MINOS~\cite{minos}, can strengthen the
present bounds.
We also review the best current limits, based on
big-bang nucleosynthesis (BBN)~\cite{Dolgov2000}.

Let us characterize the mixing of the heavy neutrinos $\nu_h$ with the
active neutrinos $\nu_a$ ($a=e,\mu,\tau$) by the corresponding element in
the mixing matrix $U$, which is the mixing matrix between the electroweak
eigenstates and the mass eigenstates.  The matrix $U$ accounts for mixing
in the neutral current (NC) interactions.  In processes mediated by the
lepton charged current (CC), the matrix $U$ enters in combination with a
unitary matrix $V$, which diagonalizes the charged lepton mass matrix.  In
principle, CC and NC interactions allow one to measure or constrain
separately the elements of $VU$ and $U$, respectively.  If one takes $V=1$,
the elements of the matrix $U$ can be interpreted as neutrino mixing
angles.   We will not make any assumptions about the matrix $V$ and will
present the constraints in full generality. 
We will assume, for
simplicity, throughout our analysis that $\nu_h$ mixes mainly with $\nu_\mu$ in the charged
current, $(VU)_{eh},(VU)_{\tau h} \sim0$, while we allow for $U_{a h} \neq
0$, $a=e,\mu,\tau$, in the neutral current.

In Section 2 we review the bounds coming from the analysis of the spectrum
of muons in pion and kaon decays.
In Section 3 we study the case of mostly-sterile neutrino decays 
into visible decay products which would be observable in a detector.
We discuss the limits on the mixing which can be obtained from 
 accelerator neutrino and Superkamiokande data and the
 prospects for strengthening such bounds in present and future
 neutrino experiments.
 In Section 4 we briefly review the bounds 
 from big bang nucleosynthesis and supernovae.
 Finally, we summarize our results in the Conclusions.

\section{Bounds on $\nu_h$ production}

The different massive neutrinos $\nu_i$, $i=1,2,3,4$, if they exist, are 
produced in
meson decays, e.g. $\pi^\pm \rightarrow \mu^\pm \nu_i$, with probabilities
that depend on the mixing in the charged current,
$VU$. 
In our analysis we will assume that the heavy sterile neutrinos
mix mainly with muon neutrinos in the charged current. 
We allow for mixing with all active neutrinos in the neutral current.
The energy spectrum of muons in such decays
would contain monochromatic lines~\cite{Shrock81}
at
\begin{equation}
T_i = ( m_\pi^2 + m_\mu^2 - 2 m_\pi m_\mu - m_{\nu_i}^2) / 2 m_\pi, 
\label{spectrummu}
\end{equation}
as long as $T_i>0$. Here $T_i$ is the muon kinetic energy; $m_\pi$,
$m_\mu$, and $m_{\nu_i}$ are the masses of pion, muon and the $i$th
neutrino mass eigenstate, respectively.  The dominant line is obtained for
nearly massless neutrinos, $\nu_{1,2,3}$, at $T_0 = 4.120 \ \MeV$.
Additional peaks in the muon energy spectrum would be present at a position
related to the mass of the heavy neutrino, Eq.~(\ref{spectrummu}), and with
a branching ratio that depends on the mixing angle.  The same is true for
muons from $K$ decays.  Searches for peaks in pion
decays~\cite{Shrock81,prodpion1,prodpion2,prodpion3,prodpion4,prodpion5,prodpion6,prodpion7,prodpion8,prodpion9} and 
in kaon decays~\cite{Asano81,Hayano82} found no
signal and set stringent bounds on $|(V U)_{\mu h}|^2$ for masses $m_h
\ltap 360 \ \MeV$.  The corresponding excluded regions from pion and kaon
decays are shown in Figs.~1 and 2 as  gray regions,
respectively.  Different lines represent limits obtained in different
experiments (see captions of Figs.~1 and 2).
%%%%%%%%%%%%%%
%

\begin{center}
\begin{figure}
\centerline{\epsfysize=8cm \epsfbox[158 1358 796 1837]{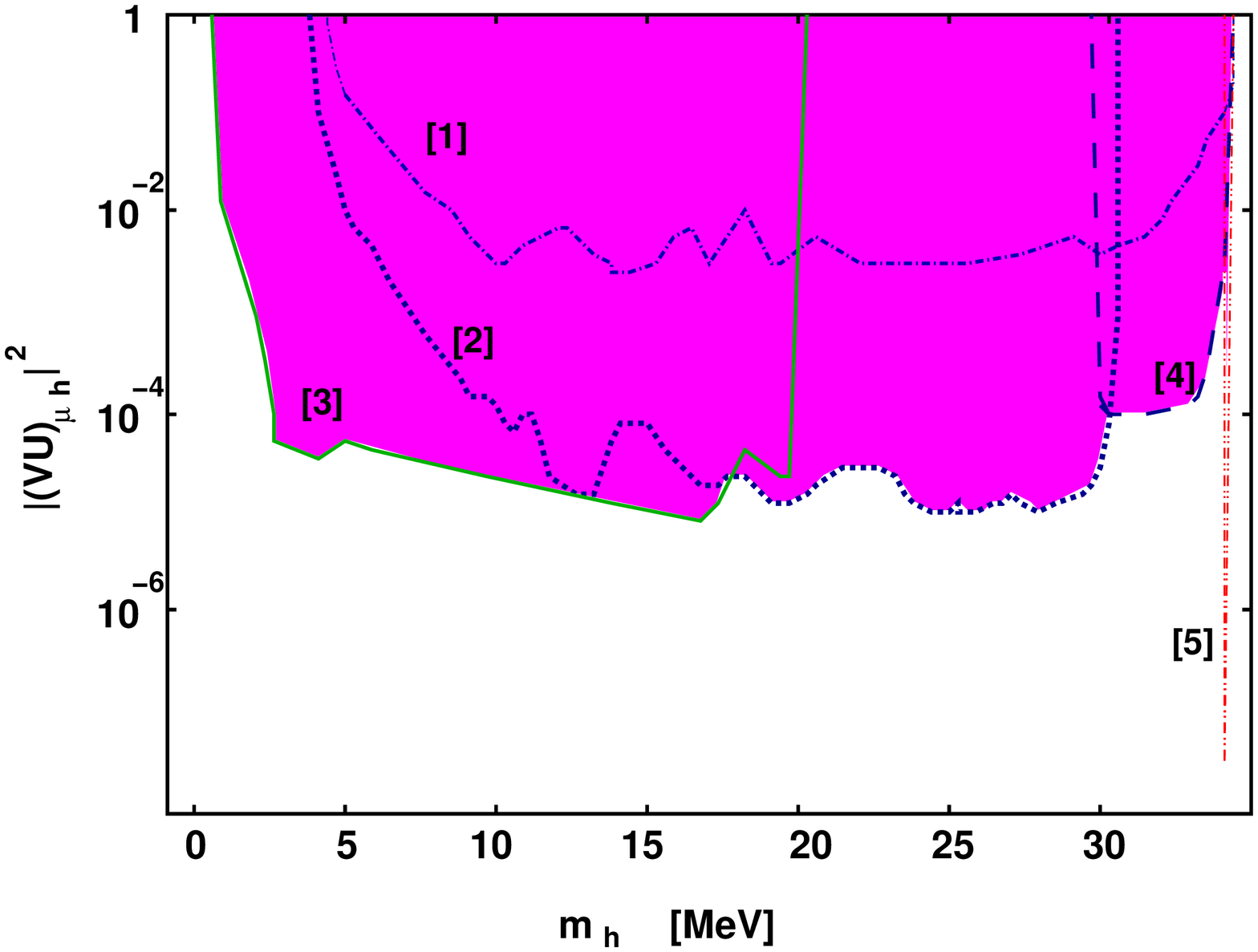}}
\caption{ The exclusion plot for $|(VU)_{\mu h}|^2$ 
based on the energy spectrum of muons in pion decays.  The
excluded region is indicated in gray color (magenta color online).  
The bounds are
taken from the analysis reported
i) in Ref.\protect\cite{prodpion2} for the dash-dotted line indicated as "[1]";
ii)  in Ref.\protect\cite{prodpion1} for the dotted line "[2]"; 
iii) in Ref.\protect\cite{prodpion4} for the solid line shown as "[3]";
iv) in Ref.\protect\cite{prodpion5} for the dashed line "[4]"; 
v) in Ref.\protect\cite{prodpion7}  for the dashed-double dotted line labelled "[5]".  
The bounds are 90\%~C.L., except for the one
marked ``[5]'', which is 95\%~C.L.
}
\label{figure:accpion}
\end{figure}
\end{center}
%%%%%%%%%%%%%%%%%
%%%%%%%%%%%%%%%
\begin{figure}
\centerline{\epsfysize=8cm 
\epsfbox[144 990 875 1508]{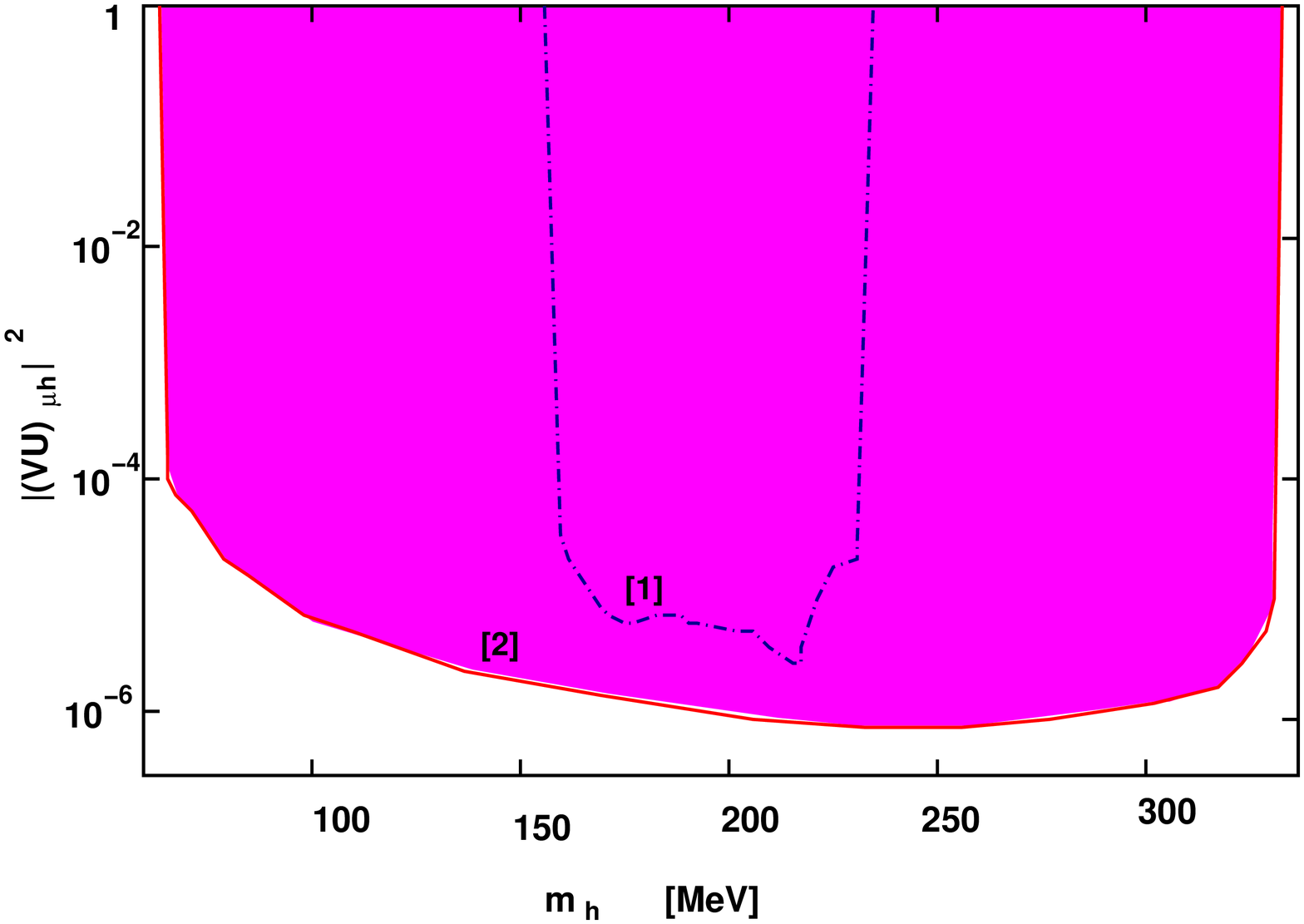}}
\caption{ The same as in Fig.~\protect\ref{figure:accpion}
but for $\nu_h$ from kaon decays.
The bounds are taken from Ref.~\protect\cite{Asano81} 
for the dash dotted line indicated as "[1]" at 2~$\sigma$
and from the data of the experiment by R.~S.~Hayano {\em et al.},
Ref.~\protect\cite{Hayano82}, 
for the solid line labelled "[2]" at 90\protect\%~C.L.
}
\label{figure:acckaon}
\end{figure}
%%%%%%%%%%%%%%%%%

\section{Bounds on $\nu_h$ decays}

 A heavy neutrino produced in $\pi^\pm$ and $K^\pm$ decays
would subsequently decay, and its decay products could be detected.  
The absence of such  detection translates into strong 
bounds on the mixing angles with active neutrinos.
In the mass range 10--390~MeV, we set a new bound 
on $|(VU)_{\mu h} U_{a h}|$, as shown in Fig.~\ref{figure:decayNC}.

A heavy neutrino mixed with active neutrinos can decay into different
channels depending on its mass.  If $1 {\ \rm MeV} \ltap m_h \ltap 105 {\
\rm MeV}$, $\nu_h$ decays via neutral currents into (i) an active neutrino and
an electron-positron pair, $\nus \rightarrow \nu_a + e^+ +e^-$, (visible
channel) and (ii) into three neutrinos, $\nus \rightarrow \nu_a + \nu_j+
\bar{\nu}_j$, (invisible channel).

In the simplest case, when sterile neutrinos directly couple only to one active neutrino
in the neutral current with mixing $U_{a h}$\footnote{
In the literature the mixing is often parametrized by a mixing angle $\theta$.
We have that $\sin^2 \theta \equiv |U_{a h}|^2$.}, the mass
of the heavy neutrino is much larger then electron mass $ m_e \ll m_h < m_{ \pi^0}$,
and we assume that in the CC current $(VU)_{eh} \sim 0$, the decay
width is given by \cite{nu_karmen}: 
\beq \Gamma_{\nus}
=\frac{1+(g_L^a)^2 +g_R^2}{768 \pi^3} G_F^2 m_{h}^5 \, |U_{a h}|^2,
\label{tau_general}
\eeq 
where $g_L^{e} = 1/2 + \sin^2 \theta_W$, $g_L^{\mu,\tau} = - 1/2 + \sin^2 \theta_W, \ g_R =
\sin^2 \theta_W$. 
If the heavy neutrino mixes with $\nu_e$ in the charged current,
the decay $\nu_h \rightarrow \nu_e e^+ e^-$ would receive
an additional contribution from CC interations.
In the absence of charged currents,
the corresponding branching ratios are equal to
$B_e =(\tilde{g}-1)/\tilde{g} \approx 0.11$ for the visible channel and 
$B_\nu = 1/\tilde{g} \approx 0.89$, for the invisible channel.
Here we have defined $\tilde{g} \equiv 
1+(g_L^a)^2 +g_R^2$.
For larger masses, 
new channels are open,
namely
$\nus \rightarrow \pi^0 \nu_\mu$  and 
$\nus \rightarrow \mu^+ \mu^- \nu_a$
which are
mediated by neutral currents, 
and $\nus \rightarrow \mu^- e^+ \nu_e$,
$\nus \rightarrow \mu^+ \mu^- \nu_\mu$
and  $\nus \rightarrow \pi^+ \mu^-$,
due to CC interactions.
For $m_h > m_\pi$, the two-body decays dominate
and typically the half-life time
is of order $\tau_h \sim (10^{-8}-10^{-9}) \  |U_{\mu h}|^{-2} \ ( |(VU)_{\mu h}|^{-2}) {\rm s}$,
for the  NC (CC) mediated processes.

\subsection{Searches for accelerator neutrino decays}

The decay length of a sterile neutrino with energy $E_h\gg m_h$ is 
$L_{\rm d}
=c \ \tau_{h}\gamma_{F}$, 
where $\gamma_F= E_h / m_h$ is the $\nu_h$ gamma factor.  
The fraction of heavy neutrinos that can reach the detector before decaying
is $\exp{(-R_{\rm cr}/L_{\rm d})}$, where $R_{\rm cr}$ is the distance from
the neutrino production site  
to the detector.
Of these neutrinos, for $h/L_d \ll 1$, 
a fraction $B_{\rm vis} \, h/L_d$ decays  
in the detector via a visible channel. 
Here  $B_{\rm vis}$ is the
branching ratio of the given decay channel and $h$ is the length of the
detector.  The number of heavy neutrino decays in the detector is then given
by (see, e.g., Ref.~\cite{Bar93})
\begin{equation}
{\sf N} = N_{\pi, K} B (M \rightarrow \mu \nus) B_{\rm vis} \frac{h}{L_d}
\Omega \epsilon
\label{ndecays}
\end{equation}
where $N_{\pi,K}$ is the number of pions and kaons, $B (M \rightarrow \mu
\nus)$ is the branching ratio of the meson decays into a muon and a heavy
neutrino, $\Omega$ and $\epsilon$ are the detector acceptance and
efficiency, respectively.
In a similar way one can consider heavy neutrinos
produced in muon decays.

%%%%%%%%%%%%%%%
%
\begin{figure}
\centerline{\epsfysize=8cm 
\epsfbox{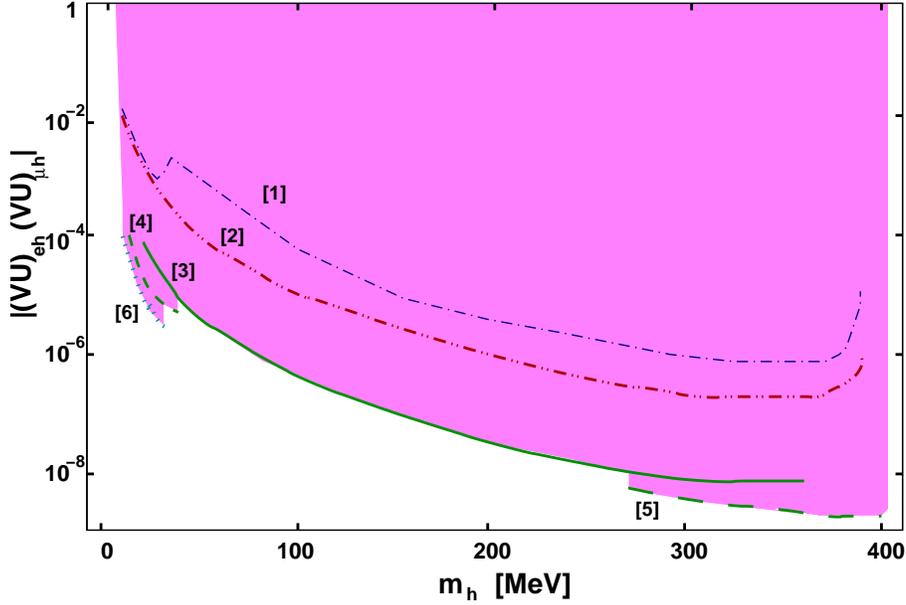}}
\caption{ The bounds on $|(VU)_{\mu h} (VU)_{e h}|$ 
versus $m_h$ obtained 
from searches of $\nu_h$ decays.
The excluded region is indicated in gray color (magenta color online).
The limits are taken
for the lines labelled as 
i) "[1]" (dashed-dotted line) from the data in Ref.~\protect\cite{Berg83} ;
ii) "[2]" (dashed-double dotted line) from the experiment in Ref.~\protect\cite{Bar93};
iii) "[3]" (solid line) from Refs.~\protect\cite{Bernardi86,Bernardi88};
iv) "[4]" and "[5]" (dashed lines) from the analysis in Ref.~\protect\cite{Bernardi88};
v) "[6]" (dotted line) from the experiment reported in Ref.~\protect\cite{Bernardi86}.
The limits are at 90\%~C.L..}
\label{figure:decayCC}
\end{figure}
%%%%%%%%%%%%%%%%%
%
%
Limits on the mixing of the heavy neutrino with $\nu_e$ and $\nu_\mu$ were
set by different experiments~\cite{Berg83,Bernardi86,Bernardi88,Bar93}.
We review them
in Fig.~\ref{figure:decayCC}.  
In the mass range $1 \ {\rm MeV} \leq m_h \leq 33.9 \ {\rm
MeV}$ the excluded region comes from heavy neutrino production in
pion decays. For higher masses,
$40 \ {\rm MeV} \ltap m_h \ltap 360\ {\rm MeV}$,
kaon decays were taken into account.
Different visible decay channels have been studied.  The channel $K^+
\rightarrow \mu^+ \nu_h \rightarrow \mu^+ (\mu^- e^+ \nu_e) + c.c.$ 
was
used to constrain the elements of the mixing matrix $VU$, 
for masses up to $\sim$260~MeV
(see Fig.~\ref{figure:decayCC}).  
For heavier masses a new decay channel is open, 
$K^+ \rightarrow \mu^+ \nu_h\rightarrow \mu^+ ( \pi^+ \mu^-)+ c.c.$,
and dominates.  

We note that in
Refs.~\cite{Berg83,Bernardi86,Bernardi88,Bar93} the NC contribution to the
decay of $\nu_h$ has not been taken into account.  They mediate the
principal decay modes both for neutrinos with $m_h < m_{\pi^0}$: ($ \nu_h
\rightarrow \nu \nu \bar{\nu}$) and with $m_h > m_{\pi^0}$: ($ \nu_h
\rightarrow \pi^0 \nu.$) However this omission does not affect the bounds
obtained in Refs.~\cite{Berg83,Bernardi86,Bernardi88,Bar93} because their visible decay channel
is always dominated by the charged current (CC) interactions.  
Since the flavor of the final neutrino is not detected in these
experiments, the channels 
$K^+ \rightarrow \mu^+ \nu_h \rightarrow
\mu^+ (e^- e^+ \nu_e)$, 
mediated by CC, and 
$K^+ \rightarrow \mu^+ \nu_h
\rightarrow \mu^+ (e^- e^+ \nu_{e,\mu,\tau})$, 
mediated by NC, are effectively
indistinguishable.  The former channel was used to constrain the $|(VU)_{eh}
(VU)_{\mu h}|$ mixing.  However, the same data can be
used for setting a bound on $|(VU)_{\mu h} U_{ah}|$,
with $a=e,\mu,\tau$, 
if one includes the contribution
due to the latter channels.

%%%%%%%%%%%%%%%

\begin{figure}
\centerline{\epsfysize=8cm 
\epsfbox{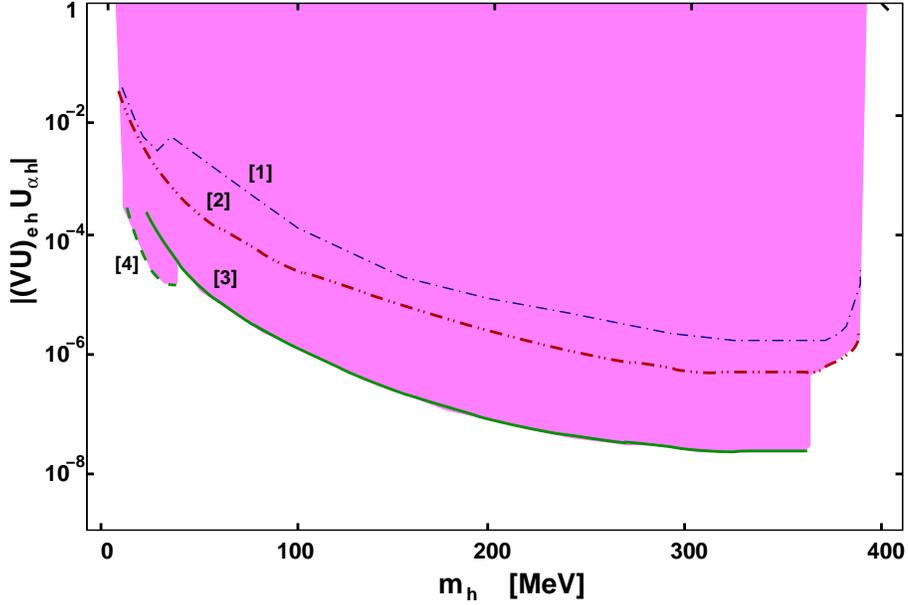}}
\caption{ The same as in Fig.~\protect\ref{figure:decayCC} 
but for the bounds on $|(VU)_{\mu h} U_{a h}|$. 
For the lines labelled as "[1]"-"[4]",
the limits are obtained from a reanalysis of the data reported in the 
references as indicated for Fig.~\protect\ref{figure:decayCC}.
}
\label{figure:decayNC}
\end{figure}
%%%%%%%%%%%%%%%%%

%%%%%%%%%%%%%%%

\begin{figure}
\centerline{\epsfysize=8cm 
\epsfbox{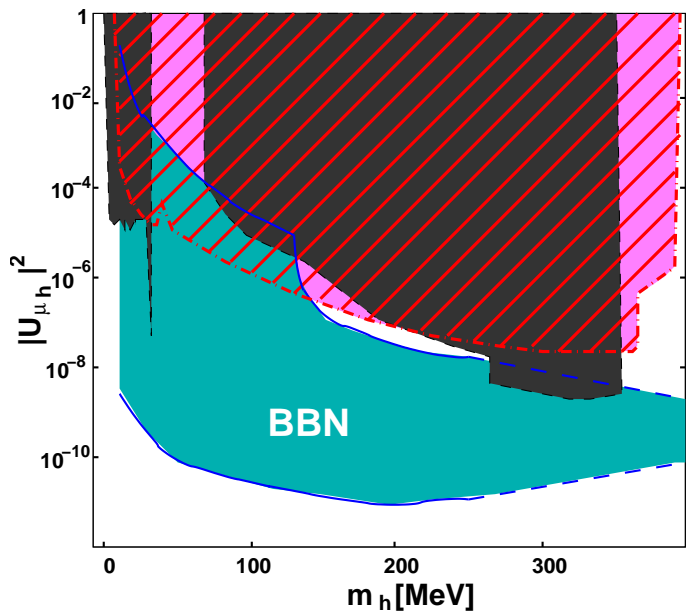}}
\caption{ The strongest bound 
(indicated as line ``[3]'' in Fig.~4 from  the
reanalysis
of data from Ref.~{\cite{Bernardi86,Bernardi88}} (diagonally hatched
region with dashed-dotted contours). The limits from
  big bang nucleosynthesis are indicated with the
  light blue-light gray region with continuous
  contours, and 
  the previous bounds from experimental searches are shown as dark gray regions with
  dashed contours. 
  For simplicity, here we take $VU = U$.
}
\label{figure:global1}
\end{figure}
%%%%%%%%%%%%%%%%%

We used the limit on $|(VU)_{eh} (VU)_{\mu h}|$
from Refs.~\cite{Berg83,Bernardi86,Bernardi88,Bar93},
corrected it by including the contribution from
the NC to the total decay width, with $(VU)_{eh} $ negligible, and the branching ratio of $K^+ \rightarrow
\mu^+ \nu_h \rightarrow \mu^+ (e^- e^+ \nu_a)$ channel, and translated it
into the new bound on $|(VU)_{\mu h} U_{ah}|$. 
These bounds have not been previously discussed in the literature
and are reported in Fig.~\ref{figure:decayNC}.
The new limits are given at the same confidence level
as the ones on $|(VU)_{eh} (VU)_{\mu h}|$.

If $V=1$, our limit on $|U_{\mu h}|^2$ 
turns out to be the strongest limit in the mass range $34 \ \MeV
\ltap m_h \ltap 200 \ \MeV$ (see Fig.~\ref{figure:global1}).  

Analogously, 
the bounds on $|U_{eh}|^2$~\cite{Bernardi86,Bernardi88} 
from $\pi^+, K^+ \rightarrow e^+ \nu_h
\rightarrow e^+ (e^+ e^- \nu_e) + c.c.$, mediated by CC interactions,
can be used to constrain
$|(VU)_{e h} U_{a h}|$
from the decays $\pi^+, K^+ \rightarrow e^+ \nu_h
\rightarrow e^+ (e^+ e^- \nu_{a}) + c.c.$,
which are induced by neutral currents.
Also these bounds are new 
but a detailed analysis is beyond the scope of the present article.

\subsection{Decays of atmospheric heavy neutrinos}

%%%%%%%%%%%%%%%

\begin{figure}
\centerline{\epsfysize=8cm 
\epsfbox{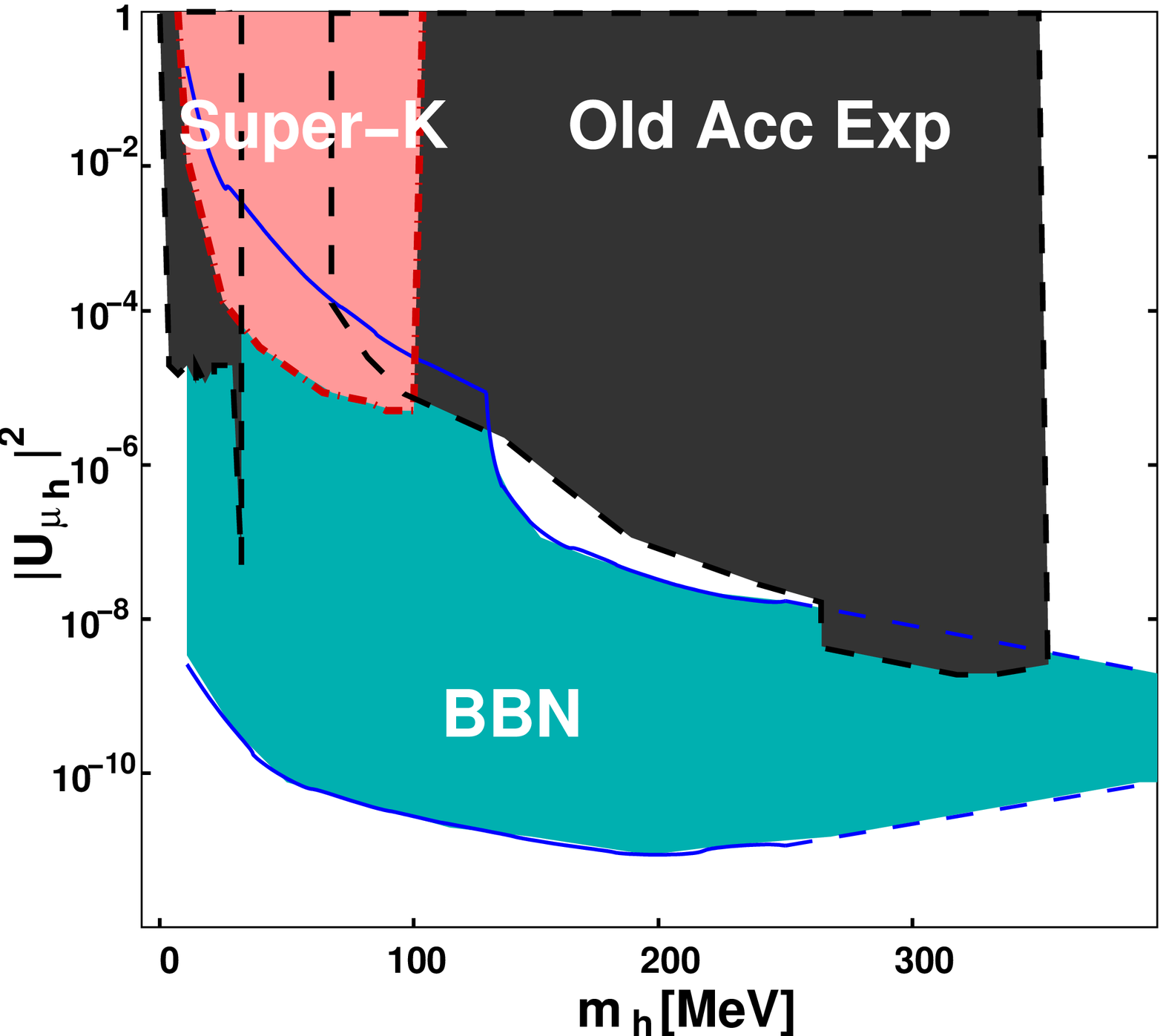}}
\caption{ The exclusion plot based on the non-observation of {$e$}-like
  events from the heavy Dirac neutrino decays in Super-Kamiokande detector at a
  rate higher than the observed $\pi^0$-like event rate~\protect\cite{SK2003} 
  (region with red dash-dotted contour). 
   Also shown are the bounds from
  big bang nucleosynthesis (light blue-light gray region with continuous
  contours), and
  from experimental searches (regions with
  dashed contours). 
  We assume $VU = U$.
}
\label{figure:global2}
\end{figure}
%%%%%%%%%%%%%%%%%

%%%%%%%%%%%%%%%%%%%
We point out that one can also set an independent limit
based on non-observation of atmospheric sterile neutrino decays by
Super-Kamiokande.  
In the following we provide a qualitative analysis
of Super-Kamiokande data in order to put a bound 
on the decays of heavy neutrinos,
if produced in the atmosphere. 
We also give an estimate for the limit on
$|U_{ah}|^2$, which we expect 
to be correct within a factor of 2--3.
We assume for simplicity that $(VU)_{\mu h}$ dominates 
in the charged current while $(VU)_{e h}\sim0$.

Heavy neutrinos can be copiously produced in the
atmosphere and can decay inside the Super-K detector generating such a high
event rate. 
The differential flux $d F_h/ dE$ of heavy sterile neutrinos
 produced in the atmosphere in pion, kaon and muon decays is related
to the active neutrino flux $F_a$ ($a=e,\mu,\tau$):
\beq \frac{dF_h(E)}{dE} = |(VU)_{\mu h}|^2 \frac{dF_\mu(E)}{dE},
\eeq 
where $dF_a(E)/dE$ is the differential flux of (active) atmospheric
neutrinos. 

The expected rate of decays, ${\sf R}$, detected in a given energy bin is
\bea 
{\sf R}
(E_1)   =    B_e \epsilon(E_1)  
 \left \{ \int_{E_1}^{E_1+\Delta E} \! \! \! \!\! dE 
\frac{dF_h}{dE} A \frac {h}{ L_{\rm d}} e^{-R_{\rm cr}/L_{\rm d}}
\right \},
\label{rate_expect} 
\eea
where $B_e$ is the branching ratio of the 
$\nu_h \rightarrow e^+ e^- \nu_a$ channel, due to NC,
 and $\epsilon(E_1)$ is
the efficiency to detect the decay products visible in the detector, $E_1$
is their average energy,
$A$ is the detector area.
This rate should not exceed the rate of events
observed by the Super-K experiment per energy bin,
${\sf R}%_{\rm decay}
(E_1)<R_{\rm event}(E_1)$.

It is clear 
that the highest number of events
should come from neutrinos with the smaller gamma-factor $\gamma_F$,
 $\gamma_F \sim2-3$ 
 (we conservatively take $\gamma_F \gtap 2$ to ensure that the production of $\nu_h$ 
 is not suppressed by threshold effects).
If the gamma factor of $\nu_h$ is small,
$\gamma_F \sim 2 - 4$, 
most of the emitted $e^-$ and $e^+$ produce in the Super-K detector
2 separate Cerenkov-light cones,
which would be interpreted as 2 $e$-like events
coming from the decay of $\pi^0\rightarrow \gamma \gamma$
($\pi^0$-like events).
We use the data reported in Ref.~\cite{SK2003}
where the invariant mass distribution for $\pi^0$-like events is shown per
energy bin of 10~MeV, for 1489.2 days and 22.5~kton of fiducial volume,
$V\sim A h$.
In the case of a three-body decay,
the invariant mass reconstructed from the energy and momentum of
$e^+$ and $e^-$ is
\begin{equation}
\label{m12}
m^2_{12}  = (p_1 + p_2)^2 = m_h^2 + m_\nu^2 -2 m_h E_\nu
\end{equation}
where $p_{1,2}$ are the four-momenta of $e^+$ and $e^-$,
and $m_\nu \ll m_h$ and $E_\nu$ are the mass and the energy
 carried away by the undetected $\nu_\mu$, 
in the reference frame of $\nu_h$.
We assume, for simplicity, that the heavy
neutrino decays in 3 relativistic particles, $\nu_a$, $e^+$ and $e^-$, each
having, on average, 1/3 of the total energy.  
The reconstructed invariant mass can be related to the
heavy neutrino mass $m^2_{12} \simeq 1/3 \ m_h^2$.

We compute the number of 2$e$-like events in each bin of $m_{12}^2$
expected from $\nu_h$ decays, with $\gamma_F \sim 2-4$.  We take into
account the energy dependence of $d F_h / d E$: the differential flux is
with good approximation constant for $E \ltap 100 \MeV$, and then it
decreases (see, e.g., Ref.~\cite{atmoflux}).  Comparing the number of
expected events from the $\nu_h$ decays with the $\pi^0$-like events,
$N_{\rm exp}$, in each invariant mass $m^2_{12}$ bin, we obtain an upper
bound on $|(VU)_{\mu h} U_{ah}|$.  We show this bound in
Fig.~\ref{figure:global2} (vertically hatched region), if $V=1$, for Dirac
sterile neutrinos which mix mainly with $\nu_\mu$.  For Majorana neutrinos
the bound is somewhat stronger because the decay rate is larger, 
as they could decay both in one channel and in its CP-conjugate.  
A more detailed analysis which includes the contribution of decays of $\nu_h$ with
$\gamma_F < 2$, threshold effects in the production of $\nu_h$ in pion and
muon decays, the zenith angle dependence and $\gamma_F$-dependence of the
suppression factor $\exp (- R_{cr}/L_d)$, should be performed but it is not
the main focus of our paper.  However we expect our conservative bound to
remain valid within a factor of a few, as it would be strengthened by the
inclusion of events with $\gamma_F<2$, the suppression $\exp (-
R_{cr}/L_d)$ is not very important for the masses and mixing parameters
considered, the number of decays with large $\gamma_F$ which could be
misinterpreted as single e-like events is suppressed by a larger $L_d$ and,
for large masses, by a smaller flux of $\nu_h$.

Most of the decay products $e^+e^-$ from 
heavy neutrinos with larger gamma factors, 
$\gamma_F \gg 3$, would produce 2 nearly overlapping
Cerenkov rings, which would be recorded as 
a single $e$-like event with twice the energy. 
Limits on the allowed range of $m_h$ and ~$|(VU)_{\mu h} U_{ah}|$
based on these events are typically of the same order of magnitude 
but are somewhat weaker than those shown in Fig.~\ref{figure:global2}.  

For masses $m_h \ltap 7-8 \ {\rm MeV}$, the Super-K threshold limits one's
ability to set a bound because only neutrinos with  $E_h \gg m_h$ would be
above the threshold.

In principle one could use the same technique to constrain 
the mixing angle for heavy sterile neutrino
with masses  up to $ m_h < m_K - m_\mu $, produced in $K$ decays.  However,
this bound would be very weak because too few neutrinos come from $K$
decays.
Our bound based on the Super-K data is a factor of few weaker than the
bound that we obtained by re-analyzing the accelerator data, as discussed
above. A much tighter bound, possibly stronger than the
present limit for masses in the 8--105~MeV range, 
could probably be
obtained if one included the directional and other information from the
Super-K data not available to us.  
The bound can be further improved in the future, after more data is
collected by Super-K or by a bigger detector, {\em e.\,g.}, Hyper-K. 

\subsection{Future accelerator experiments}

Present and future accelerator neutrino experiments 
KEK to Kamioka (K2K)~\cite{k2k}, MiniBooNE~\cite{miniboone}
 and MINOS~\cite{minos},
have the possibility of setting new, possibly stronger,
bounds on the mixing parameters.   % $U_{e h}$, $U_{\mu h}$, $U_{\tau h}$.  
Neutrinos are produced in the decay
of pions, muons and kaons. 
The number of proton on target is very high:
$10^{20}$ for K2K and $10^{21}$ for MiniBooNE and MINOS.
The average neutrino energies are around 1~GeV for K2K and MiniBooNE
and 2.5-3~GeV for MINOS.
If heavy neutrinos are produced,
they would travel toward the detectors
where the $e^{\pm}$, $\mu^\pm$, $\pi^0$ and $\pi^\pm$ 
generated in their decays would be detected.
In MiniBooNE~\cite{miniboone}, the detector 
is of length~$\sim 10$~m,
and collects the Cerenkov 
and scintillation lights.
Both K2K and MINOS have a near and 
a far detector which can both be used
to search for heavy neutrino decays.
The K2K long-baseline experiment~\cite{k2k} 
uses a near 1 Kton water Cherenkov detector
and as a far detector Super-K.
The MINOS~\cite{minos} experiment
has a near and a far iron-scintillator detectors
of length $h\sim 20-30$~m, respectively.
For the far detector timing can be used 
to reduce the backgrounds.
The neutrino beam is pulsed
with a typical duration of the spill of few $\mu$s.
The data read-out of the signal in the detector
is done in a time-window around the beam spill
of 1.5~and 10-20~$\mu$s for K2K and MINOS, respectively.
The heavy neutrinos will take longer to reach 
the detector, typically few to tens of $\mu$s
after the light neutrinos.
Therefore a sizable fraction, if not most,
of the heavy neutrino decay signal
could be time-separated from the
background given by the accelerator $\nu_e$ and $\nu_\mu$ interactions.
Directionality can be used to disentangle
a $\nu_h$ signature from active atmospheric neutrinos.

As discussed previously, depending on their masses 
and on the mixing with the active neutrinos,
the heavy neutrinos decay mainly
in $\nu_h \rightarrow 3 \nu_a$, $\nu_h \rightarrow e^+ e^- \nu_a$,
$\nu_h \rightarrow \mu^\pm e^\mp \nu_\mu$,
and for $m_h > m_\pi$ in $\pi^0 \nu_a$, $\pi^\pm e^\mp$, $\pi^\pm \mu^\mp$.
The CC induced decays would allow to constrain the product
$|(VU)_{e h} (VU)_{\mu h}|$,
while the NC processes $|(VU)_{\mu h} U_{a h}|$. 
The flux contains $\nu_h$ with different $\gamma_F$,
which would give different signatures in the detector.
A detailed analysis which takes into account
the heavy neutrino spectrum,
the different decay channels and their signals in the detector,
the relative backgrounds
needs to be performed in order to achieve a good evaluation
of the sensitivities for these experiments.
Nevertheless we can provide an order of magnitude estimate
for the limits which can be reached on the mixing parameters.
Searching for 2 $e$-like, single $e$-like, $\mu$-like and pion events,
K2K, MiniBooNE and MINOS should be able to reach sensitivities 
as good as a few$\times 10^{-7}$ for $m_h \sim 100~\MeV$.  The sensitivity 
would be of the order of a few$\times 10^{-9}$ and a few$\times 10^{-10}$
for heavy neutrino masses $m_h \sim 200~\MeV$ and $300~\MeV$, respectively.

In contrast with the other experiments, MINOS has a
very good discrimination of $\mu^+$ and $\mu^-$.
Neutrinos with $m_h \gtap 210 \ \MeV$
would decay in this channel with a branching ratio which is typically $\sim 10^{-3}$.
The decay $\nu_h \rightarrow \mu^+ \mu^- \nu_\mu$ 
receives a contribution from CC interactions
and therefore allow to constrain not only $|(VU)_{\mu h} U_{a h}|$
but also $|(VU)_{\mu h}|^2$.
As no significant background is expected for this decay mode,
this would be a clear signature of
the existence of heavy neutrinos.  The
number of events would allow to establish the value of the relevant mixing
angle and the reconstructed invariant mass the value of $m_h$.

We note that a possible future detection of the peak in the invariant mass
distribution for 2 $e$-like events from $\nu_h \rightarrow e^+ e^- \nu_a$,
after the contribution due to
$\pi^0$-decays is subtracted, 
for 2 $\mu$-like events due to $\nu_h \rightarrow \mu^+ \mu^- \nu_a$
and for $\pi^\pm$ and $e^\mp$ ($\mu^\mp$) produced in the decays
$\nu_h \rightarrow \pi^\pm e^\mp (\mu^\mp)$ decays,
would signal the existence of a sterile
neutrino and would allow a measurement of its mass and mixing.
If the different decay channels can be well distinguished,
a comparison of the observed branching ratios
with the theoretically predicted ones 
would allow to confirm the hypothesis
of heavy neutrino decays as the origin of the observed signal.
Furthermore,
if from  indipendent measurements we have that $|(VU)_{\mu h} U_{\mu h}|
\neq|(VU)_{\mu h}|^2 $,
it would be possible to establish that $V \neq 1$ and test
the existence of mixing in the charged lepton sector.

As already noticed in the previous discussion,
the data from the MiniBooNE and MINOS experiments
can be used to constrain also the combinations
$|(VU)_{eh}|^2$, $|(VU)_{e h} (VU)_{\mu h}|$, 
$|(VU)_{e h} U_{a h}|$,
once the proper decay channels and branching ratios are taken into account.
Also for these mixing terms, we expect a strengthening
of the present bounds.

\section{Bounds from cosmology and astrophysics}

Heavy sterile neutrinos in the MeV mass range
would be produced in the Early Universe and subsequently
decay affecting the predictions of Big-Bang Nucleosynthesis (BBN)
for the abundance of light elements and in particular 
of $\mbox{}^4$He (see, e.g., Ref.~\cite{Dolgov2000}).
The main effect would be to increase the energy density,
leading to a faster expansion of the Universe
and to a ealier freeze out of the $n/p$-ratio.
In addition, the decay of $\nu_h$ into light neutrinos,
in particular, $\nu_e$, would modify their spectrum
and the equilibrium of the $n-p$ reactions.
We report in Figs.~5 and 6, the bounds on the masses and mixing angle 
of the heavy sterile neutrino,
which can be obtained from big bang nucleosynthesis.
Note that in Ref.~\cite{Dolgov2000} the BBN bounds were derived 
up to $m_h$~200 MeV. It was shown that in the region 140--200 MeV 
the dominant decay chanel is into pions. We extrapolate 
those bounds above the QCD phase transition up to 400 MeV 
(dashed border of the BBN excluded region) taking into account
the change in relativistic degrees of freedom, which softens
the BBN bounds.
In order to produce a more careful
BBN bound for $m_h > 200$~MeV a detailed analysis similar to 
the one in Ref.~\cite{Dolgov2000} 
should be performed but is beyond the scope of the present study.
In Figs.~\ref{figure:global1} and \ref{figure:global2} we indicate our
estimated BBN bound above 200~MeV
with dashed lines to underline that these limits should be considered
valid within factors of a few.

In principle, SN1987A could
be used to exclude sterile neutrinos with mixing angles $10^{-7} \lsim
\umu \lsim 10^{-2}$ and masses $m_s \lsim T_{\rm core}$, where $
T_{\rm core}=30-80$~MeV is the core temperature of the neutron star at
about 0.1 second after the onset of the supernova
explosion~\cite{Dolgov2000}.  
For masses in excess of $ T_{\rm core}$, the
production of sterile neutrinos is suppressed by the Boltzmann factor.  The
emission of sterile neutrinos from the core depends on the pattern of their
mixing with active neutrinos, and in some cases the emission history can be
very complicated~\cite{fuller}.  Given the uncertainty in the SN models, 
one cannot set a reliable bound
outside the range already excluded by the combined BBN and new accelerator neutrino bounds
shown in Fig.~\ref{figure:global1}.

\section{Conclusions}

We have considered heavy sterile neutrinos
mixed mainly with muon neutrinos in the charged current.
These neutrinos would be produced in
pion and kaon decays due to their mixing with 
$\nu_\mu$.
They would subsequently
decay into standard model particles,
e.g. neutrinos, electrons and positrons.
The non-observation of the decay products 
in past dedicated experiments
has sets some stringent bounds on
the mixing angle between heavy and 
active neutrinos.
We have reanalyzed the existing 
accelerator data and set  new bounds 
on heavy neutrinos with masses  in the range 
$10 \ \MeV \ltap m_h \ltap 390 \ \MeV$.
In addition, we have used the Super-Kamiokande data to set a new
independent bound on the
mixing of a sterile neutrino with mass in the 8-105 MeV 
range.  
We have also discussed the potential of future experiments,
as K2K, MiniBooNE and MINOS, in improving the present limits.

\acknowledgments{
The authors thank B.~C.~Choudhary, R.~Saakyan and R.~Shrock for very
fruitful discussions.  We also thank G.~Fuller, S.~Nussinov,
S.~Palomares-Ruiz, and G.~Raffelt for helpful comments.  This
work was supported in part by the DOE Grant DE-FG03-91ER40662 and the NASA
ATP grant NAG5-13399.
}

\end{document}